\def\Fint{\rlap{$\Biggl\rfloor$}\Biggl\lceil}
\def\comp{{\rm C}\llap{\vrule height7.1pt width1pt depth-.4pt\phantom t}}
\def\Box{\kern1pt\vbox{\hrule height 1.2pt\hbox{\vrule width 1.2pt\hskip 3pt
 \vbox{\vskip 6pt}\hskip 3pt\vrule width 0.6pt}\hrule height 0.6pt}\kern1pt}
\def\gtwid{\mathrel{\raise.3ex\hbox{$>$\kern-.75em\lower1ex\hbox{$\sim$}}}}
\def\ltwid{\mathrel{\raise.3ex\hbox{$<$\kern-.75em\lower1ex\hbox{$\sim$}}}}
\def\Box{\kern1pt\vbox{\hrule height 1.2pt\hbox{\vrule width 1.2pt\hskip 3pt
 \vbox{\vskip 6pt}\hskip 3pt\vrule width 0.6pt}\hrule height 0.6pt}\kern1pt}
\documentstyle[epsfig,12pt]{article}

\begin{document}
\begin{titlepage}
\begin{flushright}
astro-ph/0109273 \\ UFIFT-HEP-00-23
\end{flushright}
\vspace{.4cm}
\begin{center}
\textbf{Back-Reaction Is For Real}
\end{center}
\begin{center}
L. R. Abramo$^{\dagger}$
\end{center}
\begin{center}
\textit{Theoretische Physik \\ Ludwig Maximilians Universit\"{a}t \\
Theresienstr. 37, \\ D-80333 M\"{u}nchen GERMANY}
\end{center}
\begin{center}
R. P. Woodard$^{\ddagger}$
\end{center}
\begin{center}
\textit{Department of Physics \\ University of Florida \\
Gainesville, FL 32611 USA}
\end{center}
\begin{center}
ABSTRACT
\end{center}
\hspace*{.5cm} We demonstrate the existence of a secular back-reaction
on inflation using a simple scalar model. The model consists of a
massless, minimally coupled scalar with a quartic self-interaction
which is a spectator to $\Lambda$-driven inflation. To avoid problems
with coincident propagators, and to make the scalars interact more
like gravitons, we impose a covariant normal ordering prescription
which has the effect of removing tadpole graphs. This version of the
theory exhibits a secular slowing at three loop order due to
interactions between virtual infrared scalars which are ripped apart
by the inflating background. The effect is quantified using an
invariant observable and all orders bounds are given. We also argue
that, although stochastic effects can have either sign, the slowing 
mechanism is superimposed upon them.
\begin{flushleft}
\end{flushleft}
\vspace{.4cm}
\begin{flushleft}
$^{\dagger}$ e-mail: abramo@theorie.physik.uni-muenchen.de \\
$^{\ddagger}$ e-mail: woodard@phys.ufl.edu
\end{flushleft}
\end{titlepage}

\section{Introduction}

The application of a force field in quantum field theory generally
rearranges virtual quanta and thereby induces currents and/or stresses
which modify the original force field. This is the phenomenon of {\it
back-reaction}. Famous examples include the response of QED to a
homogeneous electric field \cite{JS} and the response of generic
matter theories to the gravitational field of a black hole \cite{SWH}.
In the former case, virtual $e^+ e^-$ pairs can acquire the energy
needed to become real by tunneling up and down the field lines. The
newly created pairs are also accelerated in the electric field, which
gives a current that reduces the original electric field. The event
horizon of a black hole also causes particle creation when one member
of a virtual pair passes out of causal contact with the other by
entering the event horizon. As the resultant Hawking radiation carries
away the black hole's mass, the surface gravity rises.

Parker \cite{LP} was the first to realize that the expansion of
spacetime can lead to the production of massless, minimally coupled
scalars. Grishchuk \cite{Grish} later showed that the same mechanism
applies to gravitons. Production of these particles is especially
efficient during inflation because virtual infrared pairs become
trapped in the Hubble flow and are ripped apart from one another.
Since the mechanism requires both effective masslessness on the scale
of inflation and the absence of conformal invariance, it is limited to
gravitons and to light, minimally coupled scalars.

Our special interest is the back-reaction from this process. We
believe that the gravitational attraction between virtual infrared
gravitons gradually builds up a restoring force that impedes further
inflation. This mechanism offers the dazzling prospect of
simultaneously resolving the (old) problem of the cosmological
constant and providing a natural model of inflation in which there
is no scalar inflaton \cite{TsWo1}. The idea is that the actual
cosmological constant is not small and that this is what caused
inflation during the early universe. Back-reaction plays the
crucial role of ending inflation.

The purpose of this paper is to establish that there is a significant
back-reaction. This might appear obvious in view of the limitless
extent of particle creation otherwise. It is easy to show that about
one infrared pair emerges per Hubble time in each Hubble volume.
Denying that there is significant back-reaction implies that an
observer can watch this go on in the space around him for an arbitrarily
long period without feeling any effect.

However, the preponderance of expert opinion is highly doubtful
about the existence of a significant back-reaction. Some are concerned
with the methodology of previous studies of back-reaction. In the pure
gravity model described above, the expectation value of the gauge-fixed
metric was computed in the presence of a state which is free graviton
vacuum at $t=0$, and the resulting invariant element was reported in
co-moving coordinates,
\begin{equation}
\left\langle \Omega \left\vert g_{\mu\nu}(t,\vec{x}) dx^{\mu} dx^{\nu}
\right\vert \Omega \right\rangle = -dt^2 + e^{2 b(t)} d\vec{x} \cdot
d\vec{x} \; .
\end{equation}
When two loop effects are included the expansion rate is,
\begin{equation}
\dot{b}(t) = H \left\{1 - \left({G \Lambda \over 3\pi}\right)^2
\left[\frac16 (Ht)^2 + O(Ht) \right] + O(G^3)\right\} \; , \label{heff}
\end{equation}
where $G$ is Newton's constant, $\Lambda$ is the cosmological constant
and $H \equiv \sqrt{\Lambda/3}$ is the Hubble constant of the locally
de Sitter background \cite{TsWo2}. Unruh has criticized the procedure
of taking the expectation value of the metric first and then forming
it into an invariant measure of the expansion rate \cite{Unruh}. Linde
has no objection to gauge fixing but he believes that expectation
values obscure important stochastic effects \cite{Linde}.

There are also concerns about the putative physical mechanism behind
screening. Some doubt the causality of gravitational interactions
between particles which have been pulled out of causal contact with
one another. Others concede the reality of such a residual interaction
but maintain that it must be redshifted into insignificance by the
inflationary expansion of spacetime. Finally, there are those who
insist that back-reaction must degenerate to a small increase in the
expansion rate because the time and space average stress-energy tensor
induced by inflationary particle creation is that of a small, positive
cosmological constant.

Our response to plausible methodological concerns is cooptation. We
believe that a physical process such a back-reaction will manifest
itself in any reasonable formalism. To address Unruh's objection we
form the metric operator into an invariant measure of the local expansion
rate {\it before} taking its expectation value \cite{AbWo1}. To address
Linde's objection we set the formalism up so that stochastic samples of
this expansion operator can be taken instead of expectation values
\cite{AbWo2}.

These changes are too complicated to be implement in the pure gravity
model but they can be carried out in a scalar analog which possesses
the same combination of attractive self-interactions between massless,
conformally noninvariant quanta. The model is essentially a massless,
minimally coupled scalar with quartic self-interaction,
\begin{eqnarray}
\lefteqn{{\cal L} = -\frac12 \partial_{\mu} \phi \partial_{\nu} \phi
g^{\mu\nu} \sqrt{-g} - \frac1{4!} \lambda \phi^4 \sqrt{-g} } \nonumber \\
& & \qquad \qquad +\; {\rm counterterms} + {\rm ordering\ corrections} \; ,
\end{eqnarray}
in a locally de Sitter background. To make this system more similar to
gravity we remove tadpole graphs through a procedure of covariant
normal ordering that preserves conservation of the stress-energy tensor. When
this is done one can apply the same formalism which was used for pure
gravity to show that the expansion rate is slowed by an amount which
eventually becomes non-perturbatively strong \cite{TsWo3},
\begin{equation}
\dot{b}(t) = H \left\{1 - {\lambda^2 G \Lambda \over 2^7 3^4 \pi^5} \left[
(Ht)^4 + O(H^3 t^3)\right] + O(\lambda^3,G^2)\right\} \; . \label{scalarH}
\end{equation}
We shall demonstrate that this result is not changed by using an invariant
operator al\'a Unruh, and that it is superimposed upon an indeterminate effect
of order $\lambda$ when one takes stochastic samples al\'a Linde.

Section 2 addresses the various non-methodological objections (causality,
redshift, and modeling the source as a homogeneous, classical fluid). Section
3 motivates our decision to order the scalar analog theory so as to subtract
off tadpoles. Section 4 explains the actual procedure for accomplishing this.
In Section 5 we show that the expectation value of the invariant expansion 
operator agrees exactly with (\ref{scalarH}). We also argue that, while a 
stochastic sample can show an effect of either sign at order $\lambda$, there
is no significant change in the order $\lambda^2$ result. Our conclusions 
comprise Section 6.

\section{Physics of inflationary back-reaction}

The purpose of this section is to review the physics of inflationary
back-reaction so as to answer the three non-methodological objections
which were summarized before. We shall also give a simple explanation
of why back-reaction induces an ever-increasing, negative vacuum energy
in perturbation theory. The aim here is not rigor --- that is supplied
by the detailed calculations of Section 5. We seek rather to explain
in simple terms why the calculations turn out as they do.

It is worthwhile recalling that the inflationary production of gravitons
(and light, minimally coupled scalars) is a straightforward consequence of
the Uncertainty Principle, the existence of a causal horizon during
inflation, and the simultaneous masslessness and absence of conformal
invariance of the quanta being produced. The Uncertainty Principle requires
all quantum degrees of freedom to possess 0-point motion. In a quantum
field theory this means that virtual quanta continuously emerge from the
vacuum. Although the fact of their emergence does not depend upon the
background geometry, what happens to them subsequently does. In particular,
if there is a causal horizon then long wave length quanta {\it emerge} out
of contact with one another and so can never recombine. For this to happen
with an appreciable amplitude the quanta must of course be massless on the
scale of the horizon. They must also be sensitive to the local geometry. For
the conformally flat backgrounds characteristic of long periods of inflation
this implies that the quanta must not possess classical conformal invariance.
That is why gravitons and light, minimally coupled scalars experience
super-adiabatic amplification during inflation whereas vector gauge bosons,
spin $1/2$ fermions, and conformally coupled scalars do not.

In considering the gravitational back-reaction from inflationary particle
production note first that there is no buildup of particle density because
the 3-volume expands as new particles are created so as to keep the density
constant. When a new pair is pulled out of the vacuum the one before it is,
on average, already in another Hubble volume. However, the gravitational
field is another thing. The created particles are highly infrared (by the
standards prevailing at the time) so they do not carry very much stress-energy,
but they do carry some. This must engender a gravitational field in the
region between them. Because gravity is attractive, this field must act to
resist the Hubble flow. This is a very small effect because gravity is a
weak interaction, even on the scales usually proposed for inflation. The
feature that permits it to become significant is the cumulative nature of
the effect. Even after a newly created pair has been pulled into distinct
Hubble volumes its gravitational field must remain behind to add with
those of subsequent pairs. If nothing else supervenes to end inflation
first the gravitational field must eventually become nonperturbatively
strong.

There is little doubt that inflationary particle production goes on
because it is the usual explanation for the primordial spectrum of
cosmological perturbations \cite{MFB,LL} whose imprint on the cosmic
microwave background has been so clearly imaged by the latest balloon
experiments \cite{CMBR}. Nor is there any real doubt that {\it some}
gravitational back-reaction accrues from the process, since the usual
theory of structure formation holds that we live in complex structures
resulting from the gravitational collapse of primordial fluctuations
over the course of ten billion years. The real issues are, whether or not
there is a gravitational response {\it during} inflation, whether or not
this response grows with time, and whether or not the response slows
inflation.

Those who argue against any response at all base themselves on causality.
The source for our process is virtual particles which rapidly fall out
of causal contact with one another. Therefore, how can they interact in
{\it any} way? The flaw in this argument is the picture it implies of
exchange forces being maintained by instantaneous interaction across some
arbitrary surface of simultaneity. Were this correct an outside observer
could not feel the gravitational attraction from a pebble after it had
entered the event horizon (in some arbitrary frame) of a black hole. In
fact exchange forces are maintained by virtual quanta which propagate
causally, and those which carry the gravitational force from an object
falling into a black hole originate in the region {\it outside} the event
horizon. The same is true of objects which fall out of causal contact
in an inflating universe: they continue to feel a gravitational force
from one another that is carried in a completely causal manner from far
back along their past light cones. The process is depicted in Fig.~1.

\begin{figure}
\centerline{\epsfxsize=0.6\textwidth\epsffile{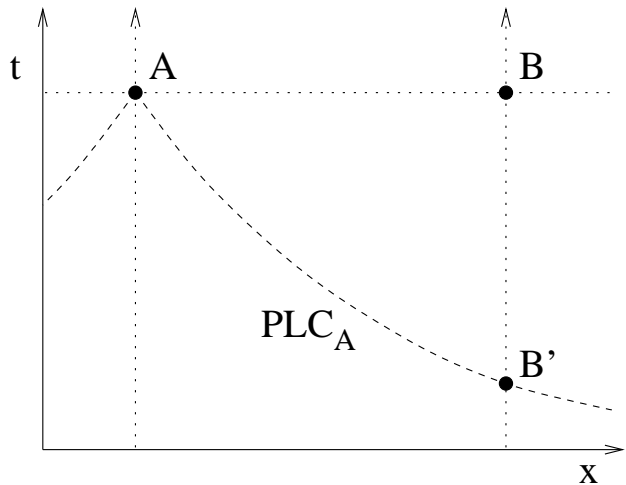}}
\caption{The worldlines of two freely falling particles $A$ and
$B$ in co-moving coordinates. The interaction $A$ feels from $B$
derives not from some arbitrary surface of simultaneity but rather
from the portion of its past light cone ($B'$ and below) in which
$B$ is visible.}
\end{figure}

Three other points deserve mention in connection with the issue of
causality. The first is that inflationary particle production
would not occur at all but for the causal horizon of an inflating
universe. Far from being something we ignore, causality is at the heart
of the effect. The second point is that the explicit perturbative
computations \cite{TsWo2,TsWo3} which support this picture were made
using the Schwinger-Keldysh formalism \cite{Schwinger}. This method
is {\it manifestly} causal: interaction vertices from outside the past
lightcone of the observation point make no contribution. The final
point is that the effects of causality are quite evident in the
factors of $Ht$ which appear in the result (\ref{heff}). They are
present, in essence, because two interaction vertices are being
integrated over the volume of the past light cone (see Fig.~2.), each
factor of which grows like $Ht$.

Those who argue against the possibility of a secular response concede
that particles continue to attract one another after having fallen out
of causal contact. But they maintain that the lines of force from this
effect must be rapidly be diluted owing to the inflationary expansion
of spacetime. Therefore the gravitational fields contributed from
particles created early during inflation should become weaker and weaker.
Were this view correct the cumulative gravitational potential would not
be the integral of a constant --- as we have argued --- but rather
some power of the ratio of past to present scale factors,
\begin{equation}
\int_0^t dt' e^{b(t') - b(t)} \sim {1 \over \dot{b}(t)} \; ,
\end{equation}
Note that we are assuming a homogeneous, isotropic and spatially flat
background geometry,
\begin{equation}
ds^2 = -dt^2 + e^{2 b(t)} d{\vec x} \cdot d\vec{x} = \Omega^2(\eta)
[ d\eta^2 + d\vec{x} \cdot d\vec{x}] \; .
\end{equation}
We have also used the slow roll approximation to evaluate the 
integral.\footnote{This means $\dot{b}^N \gg \vert b^{(N)} \vert$ and also
that $e^{b(t)} \gg 1$ for $Ht \gg 1$.}

There is no question that the argument given above applies for a
conformally invariant force field such as electromagnetism. But
gravity is not conformally invariant, and the gravitational response
from early perturbations approaches a constant rather than redshifting
to zero \cite{BG}. Because it illustrates the crucial distinction
between gravitation and electromagnetism we will make the argument
in detail. Consider an electromagnetic analog of the particle
creation process in which a homogeneous and temporally constant
current density $J^i(t,\vec{x}) = j \delta^i_{~3}$ is produced at each
point in an inflating spacetime. In flat space this would engender a
linearly growing electric field of the form, $E^i(t,\vec{x}) = - j t
\delta^i_3$, whose interpretation would be the superposition over time
of a constant electric displacement.

Things work a little differently during inflation. The curved space
Max\-well equations are,
\begin{equation}
\partial_{\nu} \left( \sqrt{-g} F^{\nu \mu}\right) = J^{\mu} \sqrt{-g} \; .
\end{equation}
For our problem only the $\mu = 3$ component is nontrivial,
\begin{equation}
-{d \over dt} \left( e^{3 b(t)} F^{30}(t)\right) = j e^{3 b(t)} \; .
\end{equation}
If we assume zero initial electric field the solution is,
\begin{equation}
E^3(t) \equiv F^{30}(t) = -j \int_0^t dt' \left(e^{b(t') - b(t)}\right)^3
\sim {-j \over 3 \dot{b}(t)} \; .
\end{equation}
The continuous current produces an electric field which approaches a
constant during inflation (when $\dot{b}$ is nearly constant). There is
no appreciable buildup from previous times because the electric field
lines redshift like $e^{-2b}$.\footnote{This exercise incidentally
illustrates the way in which sources can produce effects even when they
are not within one Hubble length on some arbitrary surface of simultaneity.
To see this simply turn the current off at a certain time and watch the
subsequent evolution of the electric field. Although it decays
exponentially it is still nonzero, despite there being no current
anywhere on the same surface of simultaneity.}

To see what happens without conformal invariance let us first translate
the electrodynamic problem to the context of a massless, conformally
coupled scalar. The Lagrangian is,
\begin{equation}
{\cal L}_{\rm MCC} = -\frac12 \partial_{\mu} \psi \partial_{\nu} \psi
g^{\mu\nu} \sqrt{-g} - \frac1{12} \psi^2 R \sqrt{-g} - J \psi \sqrt{-g} \; .
\end{equation}
Specializing its field equation to a constant source $J(x) = j$ in a
homogeneous and isotropic geometry gives,
\begin{equation}
{1 \over \sqrt{-g}} \partial_{\mu} \left( \sqrt{-g} g^{\mu\nu}
\partial_{\nu} \psi\right) - \frac16 R \psi = J \longrightarrow - e^{-2b}
{d \over dt} \left( e^b {d \over dt} e^b \psi\right) = j \; .
\end{equation}
Assuming no initial field the solution is,
\begin{equation}
\psi(t) = -j e^{-b(t)} \int_0^t dt' e^{-b(t')} \int_0^{t'} dt'' e^{2b(t'')}
\sim {-j \over 2 \dot{b}^2(t)} \; .
\end{equation}
Just as in the analogous electrodynamic problem, the conformally coupled
scalar approaches a constant during inflation because previous
contributions to it are redshifted.

We can understand what happens when conformal invariance is absent by
comparing the response to the same source from a massless, {\it minimally}
coupled scalar. Its Lagrangian is,
\begin{equation}
{\cal L}_{\rm MMC} = -\frac12 \partial_{\mu} \phi \partial_{\nu} \phi
g^{\mu\nu} \sqrt{-g} - J \phi \sqrt{-g} \; .
\end{equation}
The same specializations as before reduce the equation of motion to,
\begin{equation}
{1 \over \sqrt{-g}} \partial_{\mu} \left( \sqrt{-g} g^{\mu\nu}
\partial_{\nu} \phi\right) = J \longrightarrow - e^{-3b}
{d \over dt} \left( e^{3b} \dot{\phi}\right) = j \; .
\end{equation}
The solution can again be expressed as a double integral, but now there
is no outer redshift factor,
\begin{equation}
\phi(t) = -j \int_0^t dt' e^{-3 b(t')} \int_0^{t'} dt'' e^{3 b(t'')}
\sim -j \int_0^t dt' {1 \over \dot{b}(t')} \; .
\end{equation}
During inflation $\dot{b}$ is nearly constant, so $\phi(t) \sim -j t/\dot{b}$.
Hence we learn that lifting conformal invariance permits early sources to
contribute on an equal footing with late ones.

The scalar comparison we have just given was far from specious. It has
long been known that, dynamical gravitons in a homogeneous and isotropic
background geometry obey the same linearized equation as massless, minimally
coupled scalars \cite{Grish}. It is also well known that a single
perturbation of the metric does not redshift away during inflation.
Instead the (synchronous gauge) perturbed metric approaches the form
\cite{BG,BMT},
\begin{equation}
g_{\mu\nu}(t,\vec{x}) dx^{\mu} dx^{\nu} \longrightarrow -dt^2 + e^{2 b(t)}
\gamma_{ij}(\vec{x}) dx^i dx^j \; .
\end{equation}
Boucher and Gibbons prove their ``cosmic baldness'' theorem by then arguing
that, since the temporally constant, spatial variation of $\gamma_{ij}(
\vec{x})$ must fall off above a certain co-moving Fourier mode, any freely
falling local observer is eventually unable to sense it. This is true for
{\it one} perturbation, but what happens when there is a mechanism, such as
the Uncertainty Principle, which generates perturbations with higher and
higher co-moving wave numbers? There is simply no escaping the fact that
the amplitude of $\gamma_{ij}$ will change more and more, and that this
change will be visible to local observers. Hence the effect is not only
real but also secular.

This brings us to the third objection of principle, that inflationary
particle production should result in a small fractional {\it increase}
in the expansion rate. The basis of this argument is modeling the
stress-energy of particle production as a homogeneous and isotropic
classical fluid. If this were correct we could compute the gravitational
back-reaction using the homogeneous and isotropic Einstein equations,
\begin{eqnarray}
3 \dot{b}^2(t) & = & 3 H^2 + 8 \pi G \rho(t) \; , \label{Ein1} \\
-2 \ddot{b}(t) - 3 \dot{b}^2(t) & = & - 3 H^2 + 8 \pi G p(t) \; . \label{Ein2}
\end{eqnarray}
It is straightforward to show that locally de Sitter inflation at Hubble
constant ${\cal H}$ results in an {\it average} energy density and pressure
of $\rho = - p = {\cal H}^4/{16 \pi^2}$ per species of massless, minimally
coupled scalar.\footnote{For a detailed derivation of this old and
well-known result, see Section 4 of \cite{TsWo4}.} Substituting $\dot{b}
= {\cal H}$ and solving the resulting quadratic equation gives the
following result for the final expansion rate,
\begin{equation}
{\cal H}^2 = {3 \pi \over 2 G} \left(1 - \sqrt{1 - \frac{4 G}{3\pi} H^2}
\right) = H^2 \left\{1 + {G \Lambda \over 9 \pi} + 2 \left({G \Lambda \over
9 \pi}\right)^2 + \dots \right\} \; . \label{steady}
\end{equation}

That something must be wrong with this argument becomes apparent from
the evident fact that it requires the expansion rate to increase. In
other words, before the steady state solution (\ref{steady}) is attained
we must have $\ddot{b} = - 4\pi G (\rho + p) > 0$. But this violates the
weak energy condition! Such a thing {\it can} happen through quantum
effects \cite{Ford}, but the physics is hopelessly wrong in this case.
A collection of particles, even gravitons, should be a drag on the
expansion of spacetime, not a super-accelerant to it.

Closer inspection reveals two flaws in the argument. First, it assumes
that the distribution of created particles is homogeneous and isotropic.
This can hardly be so when the density of these quanta is about one per
Hubble volume. Gravitons move at the speed of light, and there must
necessarily be a {\it direction} associated with this motion. The one
infrared graviton in any single Hubble volume cannot give rise to a
stress-energy tensor which is even approximately isotropic. Pretending 
otherwise risks the same sort of mistake as using the zero average 
charge density to describe the dielectric properties of a ponderable 
medium which actually consists of an enormous number of positive and 
negative charges.

The second flaw is ignoring quantum correlations. Fig.~2 presents two
diagrams that contribute to the quantum gravitational back-reaction
on an inflating universe. Each has two loops and contributes at order
$(G \Lambda)^2$. Yet the diagram on the left gives a constant that can
be absorbed into renormalizing the cosmological constant\footnote{In
fact it {\it must} be so absorbed in order for the initial condition of
inflation with Hubble constant $H$ to be satisfied.} whereas the one on
the right contributes to the secular slowing of the Hubble constant
in expression (\ref{heff}) \cite{TsWo2}. The key distinction between
the two graphs is that the virtual gravitons on the right are correlated
when they interact whereas the ones on the left are not.

\begin{figure}
\centerline{\epsfxsize=0.6\textwidth\epsffile{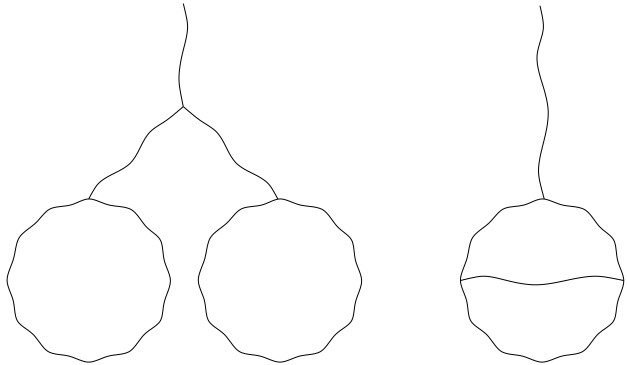}}
\caption{Two loop contributions to the background geometry from
gravitons. The lefthand diagram is a negligible constant because the
gravitons do not interact in a correlated fashion. The righthand one
slows expansion by a fractional amount which grows like $(G \Lambda)^2
(Ht)^2$ because the gravitons are correlated when they interact.}
\end{figure}

It is easy to understand why correlation can make so much difference.
Consider the traveling wave pulses depicted in Fig.~3. If the stretched
string upon which they move is exactly linear then the packets pass
through one another with no effect, however we can imagine that there is
a nonlinear interaction whose contribution to the potential energy of
the string goes like the fourth power of the displacement. Then it is
clear that pulses of the same sign and amplitude $A$ repel one another
because the superposed amplitude of $2A$ gives a potential energy of
$+16 A^4$, which is larger than the $2 A^4$ when they are far apart. On
the other hand, pulses with the opposite sign attract because their
amplitude together is zero, which means zero potential as compared with
the same $2 A^4$ when far apart. Pulses of the opposite sign are a
reasonable way (with classical string!) of representing
particle-anti-particle pairs. Ignoring their affinity for one another
--- which is what using the average stress-energy does --- is a terrible
approximation. It doesn't even get the sign right!

\begin{figure}
\centerline{\epsfxsize=0.6\textwidth\epsffile{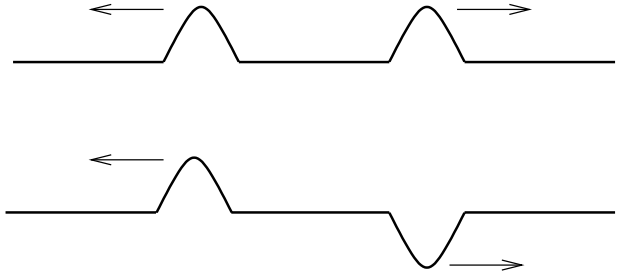}}
\caption{A simple illustration of how correlation can change an
interaction. The wave packets on the top have the same sign, so
were there a fourth order coupling they would repel one another.
The wave packets on the bottom have opposite signs. With the same
coupling they attract one another.}
\end{figure}

In considering the effects of anisotropies and correlations it is important
to distinguish between the bare stress-energy of created gravitons --- just
the $\hbar \omega$ per particle --- and the stress-energy they develop
through gravitational interaction with one another. The former is highly
anisotropic and also highly correlated, which is why one doesn't even
get the right sign by using its average value to compute the gravitational
interaction energy. What the various elements of such a source must actually
do is to attract one another, whereas a precisely homogeneous and isotropic
source causes spacetime to expand. However, the gravitational interaction
at any point is the result of superposing the gravitational fields of all
pairs that have been created in the past light cone of that point. There
are no significant quantum correlations between different pairs, and their
individual anisotropies tend to average out over a long period of inflation.
So it should be valid to infer the effect on inflation by using
(\ref{Ein1}-\ref{Ein2}), provided it is the energy density and pressure of
gravitational interaction that are substituted, and provided that these
quantities are not themselves computed by using the average of the bare
stress-energy in the homogeneous and isotropic Einstein equations.

Of course the correct way of inferring the stress-energy of gravitational
interaction is from the operator equations of quantum general relativity,
and precisely that was done to obtain (\ref{heff}) \cite{TsWo2}. However,
we can understand the result by simply forcing the average bare energy
density of created particles to attract itself as we know that the actual,
anisotropic and quantum correlated source does. Much of the essential
physics can even be captured using Newtonian gravity to estimate the
interaction energy, provided the pressure is assumed to follow from
conservation.

Consider locally de Sitter inflation on a manifold whose spatial section 
is a 3-torus. If the physical radius of the universe is initially $H^{-1}$ 
then its value at co-moving time $t$ is,
\begin{equation}
r(t) \sim H^{-1} e^{H t} \; .
\end{equation}
As mentioned before, the average of the bare energy density of inflationally
produced infrared gravitons is,
\begin{equation}
\rho_{\rm IR} \sim H^4 \; .
\end{equation}
This is insignificant compared with the energy density in the cosmological
constant ($\sim H^2/G$), and $\rho_{\rm IR}$ is in any case positive.
However, the gravitational interaction energy is negative, and it can be
enormous if there is contact between a large enough fraction of the total
mass of infrared gravitons,
\begin{equation}
M(t) \sim r^3(t) \rho_{\rm IR} \sim H e^{3 H t} \; .
\end{equation}
For example, if $M(t)$ was {\it all} in contact with itself the Newtonian
interaction energy would be,
\begin{equation}
- {G M^2(t) \over r(t)} \sim -G H^3 e^{5 H t} \; ,
\end{equation}
Dividing by the 3-volume gives a density of about $-G H^6 e^{2 H t}$, which
rapidly becomes enormous.

Of course this ignores causality. Most of the infrared gravitons needed to
maintain $\rho_{\rm IR}$ are produced out of causal contact with one another
in different Hubble volumes. The ones in gravitational interaction are those
produced within the same Hubble volume. Since the number of Hubble volumes
grows like $e^{3 H t}$, the rate at which mass is produced within a single
Hubble volume is,
\begin{equation}
{d M_1 \over dt} \sim H^2 \; .
\end{equation}
Although most of the newly produced gravitons soon leave the Hubble volume,
their gravitational potentials must remain, just as an outside observer
continues to feel the gravity of particles that fall into a black hole. The
rate at which the Newtonian potential accumulates is therefore,
\begin{equation}
{d \Phi_1 \over dt} \sim - {G \over H^{-1}} {d M_1 \over dt} \sim -G H^3 \; .
\end{equation}
Hence the Newtonian gravitational interaction energy density is,
\begin{equation}
\rho(t) \sim \rho_{\rm IR} \Phi_1(t) \sim -G H^6 H t \sim -(G H^2)^2 \cdot
H t \cdot {H^2 \over G} \; . \label{rho}
\end{equation}

Although this model is very crude it does act in the right sense --- to slow
inflation --- and it is secular. The dependence upon coupling constants is
also correct --- $(G \Lambda)^2$, characteristic of a 2-loop process like the
quantum gravitational result (\ref{heff}). This can be understood from the
fact that one must go to one loop order to see particle production, whereas
the interactions between these particles --- which is the source of the
effect --- requires another order in perturbation theory.

One other thing that this model gets right is the equation of state,
if we also assume the relativistic form of stress-energy conservation.
When the number of e-foldings $Ht$ becomes large, the fractional rate of
change of the gravitational interaction energy is negligible compared
with the expansion rate,
\begin{equation}
\vert {\dot \rho}(t) \vert \ll H \vert \rho(t) \vert \; . \label{slow}
\end{equation}
It follows from energy conservation,
\begin{equation}
{\dot \rho}(t) = -3 H \Bigl(\rho(t) + p(t)\Bigr) \; ,
\end{equation}
that the induced pressure must be nearly opposite to the energy density. In
other words, back-reaction induces negative vacuum energy. Since the key
requirement is slow accumulation in the sense of relation (\ref{slow}), the
equation of state is really a consequence of the fact that gravity is a
weak interaction on the scale of inflation.

Before closing the section we should also comment that one does not
require a complete solution of quantum gravity in order to study an
infrared process such as this. As long as spurious time dependence is
not injected through the ultraviolet regularization, the late time
back-reaction is dominated by ultraviolet finite, nonlocal terms
whose form is entirely controlled by the low energy limiting theory.
This theory must be general relativity,
\begin{equation}
{\cal L} = {1 \over 16 \pi G} (R - 2 \Lambda) \sqrt{-g} \; ,
\end{equation}
with the possible addition of some light scalars. Here ``light''
means massless with respect to $H \equiv \sqrt{\Lambda/3}$. No other
quanta can contribute effectively in this regime.

It is worth commenting that infrared phenomena can always be studied
using the low energy effective theory. This is why Bloch and Nordsieck
\cite{FBHN} were able to resolve the infrared problem of QED before the
theory's renormalizability was suspected. It is also why Weinberg \cite{SW}
was able to achieve a similar resolution for $\Lambda = 0$ quantum
general relativity. And it is why Feinberg and Sucher \cite{GFJS} were able
to compute the long range force due to neutrino exchange using Fermi
theory. More recently Donoghue \cite{JFD} has been working along the same
lines for $\Lambda = 0$ quantum gravity.

\section{Problems with coincident propagators}

As discussed in the previous section, the effect we seek to study has
two essential features. The first is that infrared virtual particles
are continually being ripped out of the vacuum and pulled apart by the
inflationary expansion of spacetime. The second crucial feature is that
these particles attract one another through a weak long range force which
gradually accumulates as more and more particles are created. The
particles we believe were actually responsible for stopping inflation
are gravitons, and the long range force through which they did it was
gravitation. However, this is not a simple theoretical setting in which
to work. It took over a year of labor to obtain the two loop result
(\ref{heff}) --- even with computer symbolic manipulation programs
\cite{TsWo2}. Before attempting to extend this feat to include invariant
observables or stochastic samples one naturally wonders whether there
is not some simpler theory we could study which manifests the same effect.

As also explained in the previous section, the prerequisites for
inflationary particle production are masslessness on the scale of
inflation and the absence of classical conformal invariance. Massless,
minimally coupled scalars have these properties --- and the lowest
order back-reaction from self-interacting scalars can be worked out on a
blackboard in about 15 minutes \cite{TsWo3}. Of course it is not natural
for scalars to possess attractive self-interactions and still remain
massless on the scale of inflation. But we do not need a {\it realistic}
model --- that is already provided by gravitation. What we seek is rather
a {\it simple} model that can be tuned to show the same physics, however
contrived and unnatural this tuning may be.

The behavior of free, massless and minimally coupled scalars in a
locally de Sitter background has been much studied \cite{BAAF,LFLP,FSW,AVLF}.
Among the curious properties of these particles are the absence of
any normalizable, de Sitter invariant states \cite{BAAF} and the
appearance of acausal infrared singularities when the Bunch-Davies
vacuum is used with infinite 3-surfaces \cite{LFLP,FSW}. The feature
that concerns us here is the assertion that taking the coincidence
limit of the propagator gives an ultraviolet divergent constant plus
a finite term which grows linearly with the co-moving time \cite{AVLF},
\begin{equation}
i\Delta(x;x) = {\rm UV} + {H^2 \over 4\pi^2} H t \; . \label{lin}
\end{equation}
Although this may not be an observable statement for free scalars
we shall argue in this section that it has three disturbing consequences
for our program of adding a $\lambda \phi^4$ self-interaction to obtain
a paradigm for infrared quantum gravity:
\begin{enumerate}
\item{The linear growth derives from the ultraviolet, not the
infrared;}
\item{Scalars develop a linearly growing mass at order $\lambda$; and,
worst,}
\item{The stress-energy violates the weak energy condition at order
$\lambda$.}
\end{enumerate}
The purpose of this section is to demonstrate these problems. We will
explain how to correct them in the next section.

We begin by giving the Lagrangian. Without ordering corrections it is,
\begin{equation}
{\cal L} = -\frac12 \partial_{\mu} \phi \partial_{\nu} \phi g^{\mu\nu}
\sqrt{-g} - \frac1{4!} \lambda \phi^4 \sqrt{-g} + {\Delta {\cal L}} \;.
\end{equation}
The various counterterms reside in ${\Delta {\cal L}}$,
\begin{eqnarray}
{\Delta {\cal L}} & = & -\frac12 {\delta m^2} \phi^2 \sqrt{-g} - \frac1{12}
{\delta \xi} (R - 4 \Lambda) \phi^2 \sqrt{-g} - {{\delta \Lambda} \over
8 \pi G} \sqrt{-g} \; , \nonumber \\
& & \qquad \qquad - \frac12 {\delta Z} \partial_{\mu} \phi \partial_{\nu}
\phi g^{\mu\nu} \sqrt{-g} - \frac1{4!} {\delta \lambda} \phi^4 \sqrt{-g}
\; .
\end{eqnarray}
The ones on the first line are of order $\lambda$ and will figure in
the considerations of this section. The ones on the second line are of
order $\lambda^2$ and will not concern us further here.

We are not quantizing gravity. The metric is a non-dynamical background
which we take to be locally de Sitter in conformal coordinates,
\begin{equation}
g_{\mu \nu}(\eta,\vec{x}) = \Omega^2(\eta) \eta_{\mu\nu} \qquad , \qquad
\Omega(\eta) = -{1 \over H \eta} = e^{Ht} \;.
\end{equation}
(Recall that $\Lambda = 3 H^2$.) To regulate the infrared problem on
the initial value surface we work on the manifold $T^3 \times R$, with the
spatial coordinates in the finite range, $-H^{-1}/2 < x^i \leq H^{-1}/2$.
We release the state in Bunch-Davies vacuum at $t = 0$, corresponding to
conformal time $\eta = -H^{-1}$. Note that the infinite future corresponds
to $\eta \rightarrow 0^-$, so the possible variation of conformal
coordinates in either space or time is at most ${\Delta x} = {\Delta \eta}
= H^{-1}$.

Because the spatial manifold is compact, wave numbers have the form,
$\vec{k} = 2 \pi H \vec{n}$, where $\vec{n}$ is a vector of integers.
Excepting for some completely irrelevant zero modes, the free field
mode sum has the form \cite{TsWo5},
\begin{equation}
\phi_I(\eta,\vec{x}) = H^3 \sum_{\vec{k}} \left\{ u(\eta,k) e^{i \vec{k}
\cdot \vec{x}} a(\vec{k}) + u^*(\eta,k) e^{-i \vec{k} \cdot \vec{x}}
a^{\dagger}(\vec{k}) \right\} \; , \label{free}
\end{equation}
where the properly normalized Bunch-Davies mode functions are,
\begin{equation}
u(\eta,k) \equiv {1 \over \sqrt{2k}} \left( \Omega^{-1} + {i H \over k}
\right) e^{-i k \eta} \; .
\end{equation}
If we define ${\Delta \eta} \equiv \eta - \eta'$ the product of the
mode function and its conjugate can be reduced to the following useful
form,
\begin{equation}
u(\eta,k) u^*(\eta',k) = {e^{-i k {\Delta \eta}} \over 2 k \Omega(\eta)
\Omega(\eta')} + {H^2 \over 2 k^3} [1 + i k {\Delta \eta} ] e^{- i k
{\Delta \eta}} \; .
\end{equation}
This is of course the combination which occurs in the propagator.

We introduce a convergence factor of $e^{-\epsilon k}$ to promote the
mode sum for the free propagator from a distribution into a well-defined
function,
\begin{eqnarray}
\lefteqn{{i\Delta}(x;x') = {H^3 \over 2 \Omega(\eta) \Omega(\eta')}
\sum_{\vec{k}} {e^{-\epsilon k} \over k} e^{-i k \vert {\Delta \eta} \vert +
i \vec{k} \cdot {\Delta \vec{x}}} } \nonumber \\
& & \hspace{2cm} + {H^5 \over 2} \sum_{\vec{k}} {e^{-\epsilon k} \over k^3}
[1 + i k \vert {\Delta \eta} \vert ] e^{-i k \vert {\Delta \eta} \vert +
i \vec{k} \cdot \vec{x}} \; . \label{msum}
\end{eqnarray}
The spatial separation vector and its norm are ${\Delta \vec{x}} \equiv
\vec{x} - \vec{x}'$ and ${\Delta x} \equiv \Vert {\Delta \vec{x}}\Vert$.
Because the range of conformal coordinates is rather small, it is an
excellent approximation to represent the mode sum (\ref{msum}) as an
integral. When this is done the result is amazingly simple \cite{TsWo5},
\begin{eqnarray}
\lefteqn{{i\Delta}(x;x') = {1 \over 4 \pi^2} {\Omega^{-1}(\eta) 
\Omega^{-1}(\eta') \over {\Delta x}^2 - (\vert {\Delta \eta} \vert - 
i \epsilon)^2}} \nonumber \\
& & - {H^2 \over 8 \pi^2} \ln\left[ H^2 \left( {\Delta x}^2 - (\vert 
{\Delta \eta} \vert - i \epsilon)^2 \right) \right] \label{prop} +
O({\Delta \eta},{\Delta x}) \; .
\end{eqnarray}
Note that the $i \epsilon$ term serves as an ultraviolet regulator. It is
in fact an exponential cutoff on the co-moving momentum.

We can now demonstrate the first problem by taking the coincidence limit. 
Setting ${\Delta \eta} = {\Delta x} = 0 $ in (\ref{prop}) gives,
\begin{equation}
{i\Delta}(x;x) = {1 \over 4\pi^2} {1 \over \Omega^2 \epsilon^2} - {H^2 \over 8
\pi^2} \ln\left[H^2 \epsilon^2 \right] \; .
\end{equation}
This is the result with a cutoff on the co-moving momentum. For an invariant
regularization we should really cut off on the {\it physical} momentum,
$k_{\rm phys} \equiv \Omega^{-1} k$. The change is easily made with the
replacement $\epsilon \rightarrow e (H \Omega)^{-1}$,
\begin{eqnarray}
{i\Delta}(x;x) & = & \left({H \over 2\pi}\right)^2 \left\{ {1 \over e^2} -
\ln(e) + \ln(\Omega) \right\} \; , \\
& \equiv & {\rm UV} + \left({H \over 2\pi}\right)^2 H t \; .
\end{eqnarray}
We are not saying this time dependence is wrong, but its origin is very
clear. It comes from a logarithmic ultraviolet divergence which is being
cut off at a time dependent point. The terrifically undesirable feature
about this is that the result can change when one changes the ultraviolet
structure of the theory, which is precisely the sector of quantum gravity
we do not know.

To see the second problem we compute the one loop scalar self mass-squared.
The diagrams are depicted in Fig.~4. A trivial application of the Feynman 
rules gives,
\begin{equation}
-i M^2_{\rm 1-loop} = -\frac{i}2 \lambda {i\Delta}(x;x) - i {\delta
m}^2 \; .
\end{equation}
The ultraviolet divergence must be absorbed with the counterterm, and
the finite part is fixed by demanding that the scalar be initially
massless. Therefore the mass counterterm is,
\begin{equation}
{\delta m}^2 = -\frac{\lambda}2 \; {\rm UV} + O(\lambda^2) \; , \label{dm}
\end{equation}
and the renormalized one loop mass squared is,
\begin{equation}
M^2_{\rm 1-loop} = {\lambda \over 8 \pi^2} \; H^2 \; Ht \; .
\end{equation}
This establishes that coincident propagators lead to a linearly
increasing mass at order $\lambda$. Of course this is undesirable
because gravitons {\it never} develop a mass.

\begin{figure}
\centerline{\epsfxsize=0.6\textwidth\epsffile{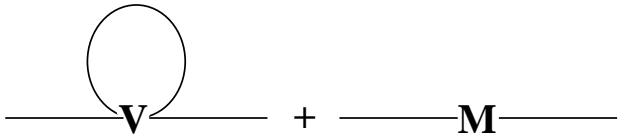}}
\caption{The scalar self-mass at order $\lambda$ {\it without} normal
ordering. $V$ denotes the 4-point vertex and $M$ stands for the
mass counterterm vertex.}
\end{figure}

To see the third and worst problem, consider the order $\lambda$ 
(two loop) corrections to the expectation value of the scalar stress-energy
tensor. They are given in Figures 5-7. The dominant contribution comes
from the diagrams in Fig.~5. Applying the Feynman rules and substituting 
for the mass renormalization (\ref{dm}) gives,
\begin{eqnarray}
T^{\rm Fig.~5}_{\mu\nu} & = & g_{\mu\nu} \left\{ -\frac{\lambda}8
\left[{i\Delta}(x;x)\right]^2 - {{\delta m}^2 \over 2} {i\Delta}(x;x)
\right\} \; , \\
& = & g_{\mu\nu} \left\{ +\frac{\lambda}8 \; {\rm UV}^2 - \left({H \over
2 \pi} \right)^4 (Ht)^2 \right\} \; .
\end{eqnarray}
Note the bizarre nature of the stress-energy tensor with this term. Inflation
actually speeds up. In fact there is a trivial interpretation for what is
happening: the Uncertainty Principle causes the scalar to wander from its
classical value of $\phi = 0$, and that engenders a positive potential
energy. It is this scalar potential energy that increases the expansion rate.

\begin{figure}
\centerline{\epsfxsize=0.6\textwidth\epsffile{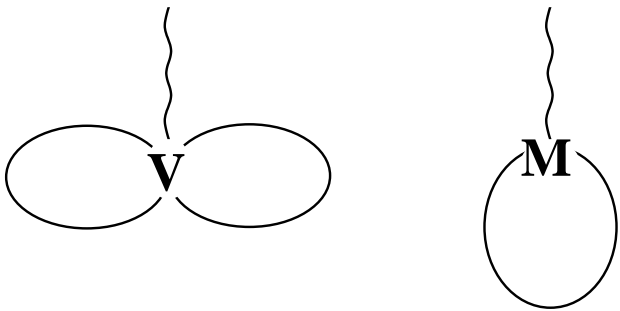}}
\caption{The dominant contributions to the scalar stress-energy tensor at 
order $\lambda$ {\it without} normal ordering. $V$ denotes the 4-point 
vertex and $M$ stands for the mass counterterm vertex.}
\end{figure}

Of course the diagrams of Fig.~5 are not the only ones which contribute to
the stress-energy tensor. Although the others cannot change the $(Ht)^2$ terms,
they do add important structure to enforce stress-energy conservation and
cancel the ultraviolet divergences.\footnote{Of particular interest is the
conformal counterterm given in the first diagram of Fig.~7. In a locally 
de Sitter background there is no distinction between $R$ and the constant,
$12 H^2$. However, the distinction {\it does} matter in computing the
stress-energy tensor and one must use the conformal counterterm to cancel
overlapping divergences which come from the diagrams of Fig.~6.}
Hence our leading order results for the induced energy density ($T_{00}
\equiv -\rho g_{00}$) and pressure ($T_{ij} \equiv p g_{ij}$),
\begin{eqnarray}
\rho & = & \frac{\lambda}8 \left({H \over 2 \pi}\right)^4 \left\{(H t)^2 
+ O(Ht)\right\} + O(\lambda^2) \; , \\
p & = & - \frac{\lambda}8 \left({H \over 2 \pi}\right)^4 \left\{(Ht)^2 +
O(Ht) \right\} + O(\lambda^2) \; ,
\end{eqnarray}
imply that their sum is,
\begin{equation}
\rho + p = -{\dot{\rho} \over 3 H} = -\frac{\lambda}8 \left({H \over 2\pi}
\right)^4 \left\{ \frac23 H t + O(1)\right\} + O(\lambda^2) \; . \label{WEC}
\end{equation}
Note that $\rho + p$ is not changed if we include the energy density and 
pressure of the bare cosmological constant through the replacements, $\rho
\longrightarrow \Lambda/{8\pi G} + \rho$ and $p \longrightarrow -\Lambda/{8
\pi G} + p$. This establishes the third and final problem with coincident 
propagators: they result in a violation of the weak energy condition at 
order $\lambda$.

\begin{figure}
\centerline{\epsfxsize=0.6\textwidth\epsffile{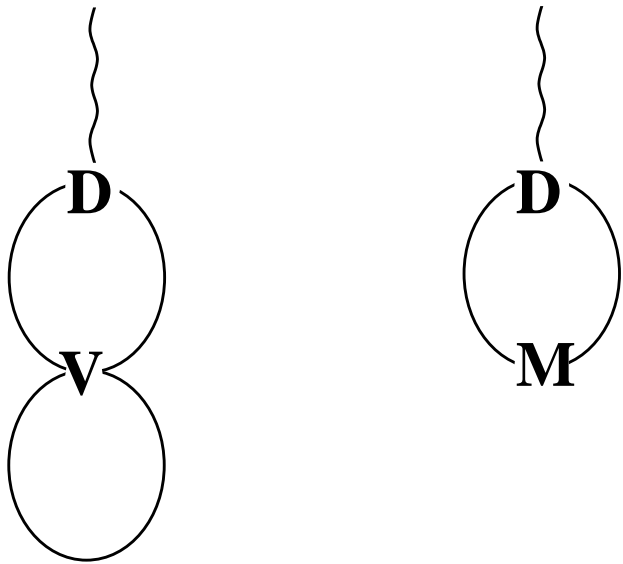}}
\caption{Contributions to the scalar stress-energy tensor at order $\lambda$
which involve the derivative vertex $D$ {\it without} normal ordering.
$V$ denotes the 4-point vertex and $M$ stands for the mass counterterm
vertex.}
\end{figure}

\begin{figure}
\centerline{\epsfxsize=0.6\textwidth\epsffile{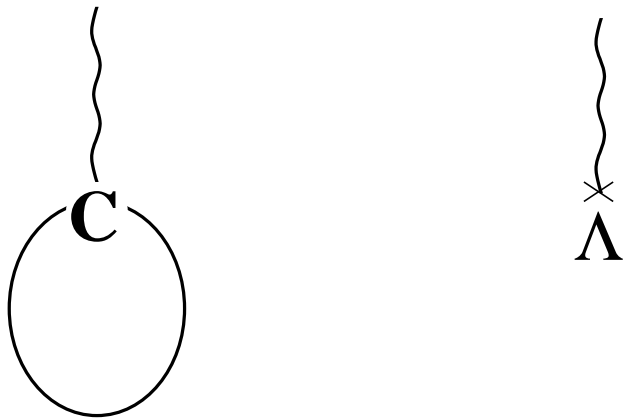}}
\caption{Contributions to the scalar stress-energy tensor at order $\lambda$
from the conformal counterterm $C$ and the counterterm for the bare
cosmological constant $\Lambda$.}
\end{figure}

\section{Covariant normal ordering}

We would like to have a theory in which the ultraviolet does not
contaminate the infrared, in which massless particles remain massless,
and in which matter is a drag on expansion rather than a super-accelerant.
The purpose of this section is to describe a consistent way in which the
scalar model can be altered to remove these undesirable features, at least
at the lowest orders in $\lambda$. We call the method, {\it covariant
normal ordering}. For simplicity we shall give the result for the original
Lagrangian, without its counterterms. Their inclusion is straightforward.

In what follows we assume that the theory has been regulated invariantly
so that the coincident propagator is a finite scalar functional of the metric. 
We define the normal-ordered product of $\phi^N$ as follows,
\begin{equation}
: \phi^N(x) : \; \equiv \sum_{k=0}^{[N/2]} {(2k-1)!! N! \over (2k)! (N-2k)!}
\left(- {i\Delta}(x;x)\right)^k \phi^{N-2k}(x) \; .
\end{equation}
Note that one can differentiate either with respect to the field or the
coincident propagator,
\begin{equation}
{\partial : \phi^N : \over \partial \phi} = N : \phi^{N-1} : \qquad , \qquad
{\partial : \phi^N : \over \partial (-i\Delta)} = \frac{N(N-1)}2 
: \phi^{N-2} : \; .
\end{equation}
The trick behind covariant normal-ordering is to implement it at the level
of the Lagrangian through the replacement: ${\cal L} \longrightarrow : {\cal 
L} :$. In this way all objects derived from the action --- such as the scalar 
stress-energy tensor and the scalar field equations --- are free of tadpoles.

Stress-energy conservation is maintained by taking account of the implicit 
metric dependence of the propagator. One can infer this, formally, from the
functional integral representation,
\begin{equation}
{i\Delta}(y;y') = \Fint [d\phi] \phi(y) \phi(y') \exp\left[-\frac{i}2 \int 
d^4x \partial_{\mu} \phi \partial_{\nu} \phi g^{\mu\nu} \sqrt{-g}\right] \; .
\label{functional}
\end{equation}
Hence the variation which gives the stress-energy tensor produces,
\begin{eqnarray}
\lefteqn{{-2 \over \sqrt{-g(x)}} {\delta {i\Delta}(y;y') \over \delta 
g^{\mu\nu}(x)} = i \left[ \delta^{\alpha}_{~(\mu} \delta^{\beta}_{~\nu)} 
- \frac12 g_{\mu\nu}(x) g^{\alpha\beta}(x)\right] } \nonumber \\
& & \times {\partial \over \partial x^{\alpha}} {\partial \over \partial 
x^{\prime \beta}} \left\{ {i\Delta}(x;x') {i\Delta}(y;y') + 2 {i\Delta}(x;y) 
{i\Delta}(x';y') \right\}_{x' = x} \; .  \label{naive}
\end{eqnarray}
However, we shall modify this scheme in two ways, one necessary and the
other highly convenient. The convenient modification is that we can drop
the first of the three terms in (\ref{naive}) because it is separately
conserved. 

The necessary modification is that one really varies the Schwinger
functional integral \cite{Jordan} to obtain the stress-energy tensor.  
This contains $+$ fields which evolve the theory forward and $-$ fields 
which evolve it back to the initial state. Although there is no mixing 
between these fields, {\it both} are minimally coupled to the same metric. 
Hence the stress-energy tensor receives contributions from both terms. This 
is necessary to make the stress-energy tensor real and to make it depend
causally upon quantities in the past lightcone of the point $x^{\mu}$ at
which it is evaluated. Two sorts of propagators result,
\begin{eqnarray}
\lefteqn{{i\Delta}_{++}(x;x') \equiv } \nonumber \\
& & {1 \over 4 \pi^2} {\Omega^{-1}(\eta) \Omega^{-1}(\eta') \over 
{\Delta x}^2 - (\vert {\Delta \eta} \vert - i \epsilon)^2} - {H^2 \over 
8 \pi^2} \ln\left[ H^2 \left( {\Delta x}^2 - (\vert {\Delta \eta} \vert 
- i \epsilon)^2 \right) \right] \; , \\
\lefteqn{{i\Delta}_{+-}(x;x') \equiv } \nonumber \\
& & {1 \over 4 \pi^2} {\Omega^{-1}(\eta) \Omega^{-1}(\eta') \over 
{\Delta x}^2 - ({\Delta \eta} + i \epsilon)^2} - {H^2 \over 8 \pi^2} 
\ln\left[ H^2 \left( {\Delta x}^2 - ({\Delta \eta} + i \epsilon)^2 
\right) \right] \; .
\end{eqnarray}
Of course the $++$ propagator is just the same as the Feynman one,
${i\Delta}(x;x')$. Note also that whereas the free kinetic operator acts 
upon ${i\Delta}_{++}(x;x')$ to give a delta function, it annihilates
${i\Delta}_{+-}(x;x')$.

With the two modifications described above the result is,
\begin{eqnarray}
\lefteqn{T_{\mu\nu}(x) = \left[\delta^{\rho}_{~\mu} \delta^{\sigma}_{~\nu}
- \frac12 g_{\mu\nu}(x) g^{\rho \sigma}(x) \right] : \partial_{\rho} \phi(x)
\partial_{\sigma} \phi(x) : - g_{\mu\nu}(x) \frac{\lambda}{4!} : \phi^4(x) :} 
\nonumber \\
& & + \frac{i \lambda}2 \left[\delta^{\rho}_{~\mu} \delta^{\sigma}_{~\nu} 
- \frac12 g_{\mu\nu}(x) g^{\rho \sigma}(x) \right] \int d^4y \sqrt{-g(y)}
: \phi^2(y) : \theta(y^0 + H^{-1}) \nonumber \\
& & \qquad \times \left[ \partial_{\rho} {i\Delta}_{++}(x;y) \partial_{\sigma}
{i\Delta}_{++}(x;y) - \partial_{\rho} {i\Delta}_{+-}(x;y) \partial_{\sigma}
{i\Delta}_{+-}(x;y) \right] \; . \quad \label{Tmn}
\end{eqnarray}
Note that $++$ and $+-$ propagators interfere destructively whenever the 
dummy variable of integration, $y^{\mu}$, strays outside the past lightcone
of the observation point $x^{\mu}$. Using the normal-ordered field equations,
\begin{equation}
{\delta : S : \over \delta \phi(x)} = \partial_{\mu} \sqrt{-g(x)} 
g^{\mu\nu}(x) \partial_{\nu} \phi(x) - \frac{\lambda}6 : \phi^3(x) : 
\sqrt{-g(x)} = 0 \; , \label{Euler}
\end{equation}
it is easy to see that the stress-energy tensor is conserved. 

Covariant normal-ordering is trivial to use: simply apply the old Feynman 
rules and then ignore any coincident propagators. Including the counterterms 
is straightforward but irrelevant because the first contributions to $\delta 
m^2$, $\delta \xi$, $\delta Z$ and $\delta \lambda$ are of order $\lambda^2$. 
Since the additional terms in the stress-energy operator consist of these 
$O(\lambda^2)$ constants times normal-ordered products of the fields, the
lowest correction to the stress-energy tensor from all except the $\delta
\Lambda$ counterterm are of order $\lambda^3$. This cannot affect the order
$\lambda^2$ effect we shall compute in the next section.

Covariant normal-ordering cannot be fundamental because it results in a 
stress-energy tensor that depends nonlocally (but causally) upon the fields.
However, it does seem to be acceptable if all we seek is a method for tuning
a simple, scalar model to make it roughly agree with what goes on in the
vastly more complicated system of quantum general relativity. In particular,
the model's stability should be enhanced, not endangered, by eliminating 
the tendency of quantum fluctuations to produce super-acceleration at the
lowest order in perturbation theory. It is also relevant to note that the 
technique reduces to ordinary normal-ordering in the flat space limit.

\section{Back-reaction}

\begin{figure}
\centerline{\epsfxsize=0.6\textwidth\epsffile{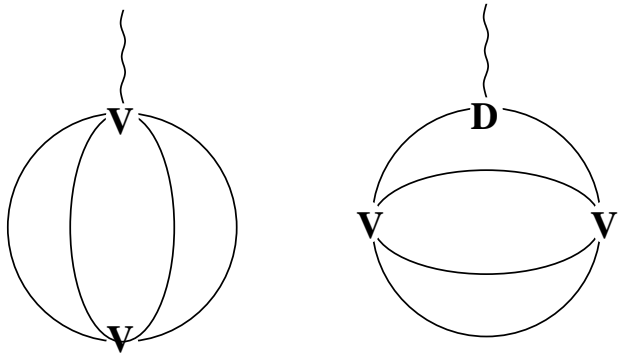}}
\caption{Contributions to the scalar stress-energy tensor at order $\lambda^2$
{\it with} covariant normal ordering. $V$ denotes the 4-point vertex and
$D$ represents the derivative vertex.}
\end{figure}

The purpose of this section is to demonstrate that the scalar model defined 
by covariant normal-ordering shows real back-reaction at the lowest order
in perturbation theory. We begin by evaluating the lowest order contribution
to the expectation value of the stress-energy tensor. This result is then 
used to compute the expectation value of the invariant expansion operator
which was defined in a previous paper \cite{AbWo1}. The conclusion is that
back-reaction slows inflation by an amount which eventually becomes 
nonperturbatively strong. All orders bounds are given for the strength of the
effect. Finally, we argue that significant back-reaction would show up as 
well if stochastic samples, rather than expectation values, had been used.

With covariant normal-ordering the expectation value of the stress-energy
tensor is much simpler than without. Because there are no coincident
propagators the lowest contribution comes at order $\lambda^2$ from the
two diagrams in Fig.~8 \cite{TsWo3}. (Of course there is a cosmological
counterterm to absorb the ultraviolet divergence.) Further, the derivatives
on the top vertex of the righthand diagram render its contribution
subdominant to the lefthand diagram in powers of $Ht$. So the dominant
contribution to the expectation value of the stress-energy tensor is 
$- g_{\mu\nu}(x)$ times,
\begin{eqnarray}
\lefteqn{\frac{\lambda}{4!} \left\langle \Omega \left\vert : \phi^4(x) : 
\right\vert \Omega \right\rangle =} \nonumber \\
& & \frac{-i \lambda^2}{4!} \int_{t'>0} d^4x' \Omega^4(\eta') \left\{ \left[ 
{i\Delta}_{++}(x;x') \right]^4 - \left[ {i\Delta}_{+-}(x;x') \right]^4 \right\}
+ O(\lambda^3) \; . \label{phi4}
\end{eqnarray}
The difference of $++$ and $+-$ propagators comes from using the 
Schwinger-Keldysh formalism \cite{Schwinger,Jordan} to compute an 
expectation value rather than an in-out amplitude. This form ensures that 
the result is real and that it depends only upon points $x^{\prime \mu}$ 
in the past lightcone of the observation point $x^{\mu}$. The lower limit 
of temporal integration at $\eta' = -H^{-1}$ (that is, $t' = 0$) derives 
from the fact that we release the state in free Bunch-Davies vacuum at 
this instant.

Although (\ref{phi4}) was computed in ref.~\cite{TsWo3} we will go over it 
in detail. Since only the logarithm term of the propagator breaks conformal 
invariance it is perhaps not surprising that the dominant secular effect 
comes from taking this term in each of the four propagators. This contribution 
is completely ultraviolet finite, and its evaluation is straightforward if 
one goes after only the largest number of temporal logarithms,
\begin{eqnarray}
\lefteqn{\frac{-i \lambda^2}{4!} \left({-H^2 \over 8 \pi^2}\right)^4 
\int_{-H^{-1}}^{\eta} d\eta' \left({-1 \over H \eta'}\right)^4 4\pi 
\int_0^{\infty} dr r^2} \nonumber \\
& & \qquad \times \left\{ \ln^4\left[ H^2 \left( r^2 - ({\Delta \eta} -i
\epsilon)^2\right) \right] - \ln^4\left[ H^2 \left( r^2 - ({\Delta \eta} +i
\epsilon)^2\right) \right] \right\} \nonumber \\
& & \rightarrow {-i \lambda^2 H^4 \over 2^{13} 3^1 \pi^7} \int_{-H^{-1}}^{\eta}
d\eta' {1 \over \eta^{\prime 4}} \int_0^{\Delta \eta} dr r^2 \; 8\pi i \;
\ln^3\left[H^2 ({\Delta \eta}^2 - r^2)\right] \; , \\
& & = {\lambda^2 H^4 \over 2^{10} 3^1 \pi^6} \int_{-H^{-1}}^{\eta} d\eta' 
{{\Delta \eta}^3 \over \eta^{\prime 4}} \int_0^1 dx x^2 \left[ 2 \ln(H 
{\Delta \eta}) + \ln(1 - x^2)\right]^3 \; , \\
& & \rightarrow {\lambda^2 H^4 \over 2^7 3^2 \pi^6} \int_{-H^{-1}}^{\eta}
d\eta' {{\Delta \eta}^3 \over \eta^{\prime 4}} \ln^3(H {\Delta \eta}) \; .
\end{eqnarray}
For large $Ht$ the biggest effect comes from the term with the most factors
of $\ln(-H\eta) = -Ht$. That the integrand contributes three such factors
follows from the expansion,
\begin{equation}
\ln(H{\Delta \eta}) = \ln(-H \eta') - \sum_{n=1}^{\infty} \frac1{n} \left({
\eta \over \eta'}\right)^n \; .
\end{equation}
An additional factor comes from performing the integration up against the 
final term in the expansion of the ratio,
\begin{equation}
{{\Delta \eta}^3 \over \eta^{\prime 4}} = {\eta^3 \over \eta^{\prime 4}} - 3
{\eta^2 \over \eta^{\prime 3}} + 3 {\eta \over \eta^{\prime 2}} - \frac1{
\eta'} \; .
\end{equation}
The final result is therefore,
\begin{equation}
\frac{\lambda}{4!} \left\langle \Omega \left\vert : \phi^4(x) : \right\vert
\Omega \right\rangle = - {\lambda^2 H^4 \over 2^9 3^2 \pi^6} \left\{ (Ht)^4
+ O(H^3 t^3) \right\} + O(\lambda^3) \; . \label{final}
\end{equation}

Three points deserve comment before we consider the effect on the invariant
expansion observable. First, there is nothing paradoxical about the negative
sign of the $(Ht)^4$ contribution to expectation value of a positive 
operator. The actual result is dominated by a positive ultraviolet divergent 
constant. It is only after the cosmological counterterm is used to subtract 
this divergence that the ultraviolet finite factor of $(Ht)^4$ dominates the
late time behavior of the scalar stress-energy tensor at order $\lambda^2$. 

Our second comment is that the negative sign of the $(Ht)^4$ term has a 
simple physical interpretation. As the inflationary expansion rips more and 
more scalars out of the vacuum their attractive self-interaction acts to 
pull them back together. This is the direct analog of the graviton effect 
we have been seeking.

Our final comment is that the full stress-energy, including the bare
cosmological constant, obeys the weak energy condition. Our leading order
result implies,
\begin{eqnarray}
{\rho(t) \over H^4} & = & {9 \over 8 \pi G \Lambda} - {\lambda^2 \over 2^9 
3^2 \pi^6} \left\{ (Ht)^4 + O(H^3 t^3) \right\} + O(\lambda^3) \; , \\
{p(t) \over H^4} & = & -{9 \over 8 \pi G \Lambda} + {\lambda^2 \over 2^9 3^2 
\pi^6} \left\{ (Ht)^4 + O(H^3 t^3)\right\} + O(\lambda^3) \; .
\end{eqnarray}
Since $G \Lambda \ll 1$, the sign of the total energy density is positive, 
even though the scalar contribution is negative. (At least for as long as 
perturbation theory remains valid.) From conservation we see that the sum of
the energy density and the pressure is positive,
\begin{equation}
\rho(t) + p(t) = {-\dot{\rho}(t) \over 3 H} = {\lambda^2 H^4 \over 2^9 3^2 
\pi^6} \left\{\frac43 (Ht)^3 + O(H^2 t^2) \right\} + O(\lambda^3) \; .
\end{equation}
This is the sense in which matter ought to act, so we conclude that covariant
normal-ordering has succeeded in making scalars behave analogously to the 
vastly more complicated graviton model.

We have so far been working in a locally de Sitter geometry. To quantify the 
back-reaction on inflation we must instead regard de Sitter as the background 
upon which perturbative corrections are superimposed,
\begin{equation}
g_{\mu\nu}(x) \equiv \Omega^2(\eta) \left[ \eta_{\mu\nu} + \kappa 
\psi_{\mu\nu}(x) \right] \qquad , \qquad \Omega(\eta) \equiv {-1 \over H \eta} 
\; .
\end{equation}
Here $\kappa^2 \equiv 16 \pi G$. The pseudo-graviton field, $\psi_{\mu 
\nu}(x)$, is determined, up to a diffeomorphism, by solving the Einstein 
equations,
\begin{equation}
R_{\mu\nu} - \frac12 R g_{\mu\nu} = -\Lambda g_{\mu\nu} + 8\pi G 
T_{\mu\nu}[g,\phi] \; . \label{Einstein}
\end{equation}
Note that this need not entail quantizing gravity. It is perfectly consistent
to suppress dynamical graviton degrees of freedom so that the pseudo-graviton
field is only an operator through its dependence upon $\phi$, and this is
what we shall do. One consequence is that $\kappa \psi_{\mu\nu}$ receives its
first nonzero contributions at order $G$. If these result in a secular 
reduction of the expansion rate --- as we will see that they do --- then 
we can establish that back-reaction is real without needing to consider 
either higher $G$ corrections to $\kappa \psi_{\mu\nu}$ or the effect of 
more than a single power of the pseudo-graviton field in the expansion 
operator. 

For the purpose of quantifying back-reaction in this system it suffices to
use the simplest of the invariant observables previously constructed for 
that purpose \cite{AbWo1}. We first define a scalar measure of the expansion
rate and then evaluate it on a geometrically fixed observation point. Our
scalar is the inverse conformal d'Alembertian acting upon a unit source,
\begin{eqnarray}
{\cal A}[g](x) & \equiv & \left({1 \over \Box_c} 1 \right)(x) \; , \\
& = & {\cal A}_0(x) + \kappa {\cal A}_1(x) + O(\kappa^2 \psi^2) \; .
\end{eqnarray}
The zeroth order term can be evaluated exactly, although we shall make the
slow roll approximation (denoted by an arrow),
\begin{eqnarray}
{\cal A}_0(x) & \equiv & -{1 \over \Omega} {1 \over \partial^2} \Omega^3\; ,\\
& = & -e^{Ht} \int_0^t dt' e^{-Ht'} \int_0^{t'} dt'' e^{2 H t''} \; , \\
& = & -{1 \over 2 H^2} \left(1 - e^{-Ht}\right)^2 \; , \\
& \longrightarrow & -{1 \over 2 H^2} \; .
\end{eqnarray}
Because this is constant, no perturbative shift in the observation point can
affect the first order result! Therefore, it does not even matter how the
observation point is geometrically determined although, for the sake of
completeness, we employ zero shift with the following clock function to fix 
surfaces of simultaneity,
\cite{AbWo1},
\begin{equation} 
{\cal N}[g](x) \equiv -\left( {1 \over 4 \Box} R\right)(x) \longrightarrow
H t + O(\kappa \psi) \; .
\end{equation}

The single graviton correction to ${\cal A}[g](x)$ is simple to evaluate in the 
slow roll approximation \cite{AbWo1},
\begin{eqnarray}
\lefteqn{\kappa {\cal A}_1(x) = -{\kappa \over \Omega} {1 \over \partial^2}
\left\{ -\psi^{\mu\nu} \partial_{\mu} \partial_{\nu} - (\psi^{\mu\nu}_{~~ ,\nu}
-\frac12 \psi^{,\mu}) \partial_{\mu} \right. } \nonumber \\
& & \hspace{5cm} \left. - \frac16 (\psi^{\mu\nu}_{~~ , \mu\nu} 
- \psi^{,\mu}_{~~ \mu})\right\} {1 \over \partial^2} \Omega^3 \; , \qquad \\
& & \longrightarrow {1 \over 24 H^4 \Omega^2(\eta)} \left( \kappa \psi^{,
\mu}_{~~\mu}(x) - \kappa \psi^{\mu\nu ,}_{~~~ \mu\nu}(x)\right) \; .
\end{eqnarray}
This particular combination of the pseudo-graviton field happens to be fixed
by the Einstein equation (\ref{Einstein}),
\begin{equation}
\kappa {\cal A}_1(x) \longrightarrow - {\pi G \over 3 H^4} g^{\mu\nu}(x) 
T_{\mu\nu}(x) \; . \label{A1}
\end{equation}
Since we need not consider corrections of higher order in $G$, the metric
in this last expression can be set to the non-dynamical background, $g_{\mu
\nu}(x) \longrightarrow \Omega^2(\eta) \eta_{\mu\nu}$. From the form of the
stress-energy tensor (\ref{Tmn}), and our previous result for the leading
contribution to its expectation value (\ref{final}), we see that the slow
roll approximation for the expectation value of the expansion operator is,
\begin{equation}
\left\langle \Omega \left\vert {\cal A}[g](x) \right\vert \Omega \right\rangle
\longrightarrow -{1 \over 2H^2} \left\{1 + {\lambda^2 G \Lambda \over 2^6 
3^4 \pi^5} \left[(Ht)^4 + O(H^3 t^3)\right] + O(\lambda^3,G^2)\right\} \; .
\label{A1exp}
\end{equation}
It follows that expectation values give the following estimate for the 
back-reacted expansion rate,
\begin{equation}
H_{\rm eff}(x) = H \left\{1 - {\lambda^2 G \Lambda \over 2^7 3^4 \pi^5} 
\left[(Ht)^4 + O(H^3 t^3)\right] + O(\lambda^3,G^2)\right\} \; .
\end{equation}
This is precisely the same result that was previously obtained by studying the
expectation value of the gauge fixed metric \cite{TsWo3}. Back-reaction is 
for real.

Although we will not compute them here, it is straightforward to estimate the
strength of higher order effects. Consider a diagram with $2N$ external scalar 
lines. At $L$ loop order the number of $\phi^4$ interaction vertices is,
\begin{equation}
V = L + N - 1 \; .
\end{equation}
Each contributes a factor of $\lambda$, so the diagram goes like $\lambda^{L
+ N - 1}$. The number of internal propagators is,
\begin{equation}
P = 2L + N - 2 \; .
\end{equation}
The largest secular effect comes, as it did for (\ref{final}), from the term
where each propagator contributes its logarithm part. Since we are computing
a Schwinger diagram, there will be $V$ cancellations between $+$ and $-$
variations, which give the $\theta$-function imaginary part of the logarithm.
However, there are also $V$ temporal integrations, each one of which can
potentially result in an extra factor of $\ln (-H \eta) = - Ht$. Hence the
strongest possible effect for the $2N$-point vertex at $L$ loop order is,
\begin{equation}
V_{2N}^L \sim \lambda^{L + N - 1} (Ht)^{2L + N - 2} \; .
\end{equation}

The stress-energy tensor corresponds to $N=0$ so the dominant contribution
at $L$ loop order is,
\begin{equation}
T_{\mu\nu}^L \sim g_{\mu\nu} H^4 \left(\lambda (Ht)^2\right)^{L-1} \; .
\end{equation}
It follows that perturbation theory breaks down at $H t \sim 1/\sqrt{\lambda}$.
Since $\lambda$ is assumed small we see that back-reaction can be studied
reliably for an enormous number of e-foldings. Note that all the higher point 
diagrams remain perturbatively weak during this entire period,
\begin{equation}
\lim_{Ht \rightarrow \lambda^{-1/2}} V^L_{2N} \sim \lambda^{N/2} \; .
\end{equation}
It should therefore be valid to use perturbation theory almost up to $Ht = 
1/\sqrt{\lambda}$.

Finally, we consider the effect of taking stochastic samples of ${\cal 
A}[g](x)$ rather than computing its expectation value. A procedure for 
implementing this perturbatively was worked out in ref. \cite{AbWo2}. What one
does is to solve the scalar field equations (\ref{Euler}) for $\phi(x)$ in 
terms of its initial value and that of its first derivative, organized as 
creation and annihilation operators on free Bunch-Davies vacuum,
\begin{equation}
\phi(x) = \phi_I(x) + \frac{\lambda}6 \int_{t' > 0} d^4x' G_{\rm ret}(x;x') : 
\phi_I^3(x') : + O(\lambda^2) \; . \label{phiexp}
\end{equation}
The free field $\phi_I(x)$ was given in expression ({\ref{free}), and the 
retarded Green's function is,
\begin{equation}
G_{\rm ret}(x;x') = - {\theta({\Delta \eta}) \over 4 \pi} \left\{ {\delta(
{\Delta \eta} - {\Delta x}) \over \Omega(\eta) \Omega(\eta') {\Delta x}}
+ H^2 \theta({\Delta \eta} - {\Delta x}) \right\} \; . \label{Gret}
\end{equation}
The order $G$ result for the expansion observable comes from substituting 
this solution into (\ref{A1}) and then assigning random $\comp$-number values 
to those creation and annihilation operators in $\phi_I(x)$ which have 
experienced horizon crossing by the observation time. 

Since the dominant contribution to the stress-energy tensor is from the 
quartic coupling we can make the replacement,
\begin{equation}
\kappa {\cal A}_1(x) \longrightarrow - {\lambda G \over 18 H^4} : \phi^4(x) :
\; .
\end{equation}
The order $\lambda$ contribution to this operator is obtained by replacing
all the fields $\phi(x)$ by the free field $\phi_I(x)$. Although this term
has zero expectation value on account of covariant normal-ordering, a
stochastic sample will generally be nonzero. We can compute its variance by 
taking the expectation value of its square,
\begin{equation}
\left\langle \Omega \left\vert \left(: \phi_I^4(x) : \right)^2 \right\vert 
\Omega \right\rangle = 24 [{i\Delta}(x;x)]^4 \; .
\end{equation}
Since ${i\Delta}(x;x)$ grows like $H^2 /{4\pi^2} Ht$, we see that the order 
$\lambda$ contributions to $\kappa {\cal A}_1(x)$ fluctuate about zero with 
standard deviation,
\begin{equation}
\sigma_{\kappa {\cal A}_1} = {1 \over 2 H^2} \left\{ {\lambda G \Lambda \over 
2^2 3^{2.5} \pi^3} \left[(Ht)^2 + O(Ht)\right] + O(\lambda^2,G^2)\right\} \; .
\end{equation}
For some samples the fluctuation reduces the expansion rate from its classical
value, for other samples the expansion rate is increased. 

It is important to realize that this order $\lambda$ effect is the same as 
the one previously studied on the nonperturbative level by Linde and 
collaborators \cite{Linde}. Although its variance does have the stated 
temporal dependence, what an actual observer sees is the effect of the
field executing a drunkard's walk. Hence the time dependence of the order 
$\lambda$ contribution to a stochastic sample of ${\cal A}[g](x)$ is not 
simple. 

The order $\lambda^2$ effect we saw from the expectation value derives from 
the term where three of the fields in $: \phi^4 :$ are the free field 
$\phi_I$ and the other field is the order $\lambda$ correction in 
(\ref{phiexp}). The nonlocal character of this term tends to wash out its 
variance \cite{AbWo2}, so stochastic samples of it are clustered tightly 
around the mean value (\ref{A1exp}). We conclude that stochastic samples of 
the expansion operator consist of the same secular slowing term we found in 
its expectation value, superimposed upon a genuinely stochastic, random walk 
at order $\lambda$. Since there is only a small probability for the drunkard's
walk to exhibit monotonic time dependence, it is possible to distinguish the 
two effects, even if a fluctuation happens to make the order $\lambda$ 
contribution larger. Therefore, back-reaction is still real in the presence 
of stochastic effects.

\section{Discussion}

In this paper we have employed a simple scalar model to demonstrate that 
there can be significant back-reaction on inflation, even when the effect is 
quantified using an invariant operator measure of the expansion rate and even 
when stochastic effects are included. The expectation value of the invariant
expansion operator gives precisely the same result that was previously 
inferred by computing the expectation value of the gauge-fixed metric
\cite{TsWo3}. The situation is more complicated when stochastic effects are 
included. The result in this case is that almost the same secular slowing is 
superimposed upon the perturbative analog of the stochastic effect previously
studied by Linde and collaborators \cite{Linde}. In both cases back-reaction
slows inflation by an amount which eventually becomes nonperturbatively
strong.

The scalar model is somewhat contrived in two ways. First, the scalar's 
natural mass is not zero but rather the scale of inflation, $\sqrt{H 
M_{\rm pl}}$. Second, the covariant normal-ordering prescription of 
Section 4 results in a subdominant contribution to the stress-energy 
tensor (\ref{Tmn}) which depends causally but nonlocally upon the scalar 
and metric fields. This term plays no role at the order we worked, but 
it is necessary to enforce conservation at higher orders. Neither of 
these features should cast doubt upon the reality of back-reaction in 
quantum gravity. In fact they were imposed upon the scalar model to 
make it more nearly resemble gravity.

In Section 2 we have also tried to answer the three objections of principle
sometimes made to a significant back-reaction: causality, redshift, and 
averaging the source. To briefly recapitulate, the gravitational 
attraction between superhorizon particle pairs derives from virtual gravitons 
which were emitted in the past, before each particle exited its partner's 
causal horizon. The same mechanism is responsible for the persistence of 
gravitational fields due to massive objects which have fallen inside the 
event horizon of a black hole. Although electromagnetic fields are redshifted 
by inflation, gravitational potentials are not. This derives from the fact 
that electromagnetism is conformally invariant whereas gravity is not. On a 
concrete level one can see it from the theta function term in the retarded 
Green's function (\ref{Gret}) which is common to minimally coupled scalars and 
dynamical gravitons. Finally, it is not valid to compute back-reaction from 
the average stress-energy of produced particles because this suppresses the 
quantum correlation between produced {\it pairs} and because the actual 
distribution of particles is not uniform. Clumps of energy density attract 
one another gravitationally whereas the perfectly uniform distribution 
which results from taking the spatial average simply increases the overall
expansion rate.

A spinoff of our work is the order $\lambda$ violation (\ref{WEC}) of the 
weak energy condition when the scalar model is {\it not} covariantly 
normal-ordered. This seems to be an analog, on cosmological scales, of 
quantum violations of the energy conditions which have been previously 
studied on much smaller scales \cite{Ford}. With observations on the 
current state of the universe not disfavoring an equation of state with 
$w < -1$ \cite{Caldwell} it is worth taking note of models that can achieve 
this.

\vskip .5cm
\centerline{\bf Acknowledgments}

We have profited from conversations on this subject with R. H. Brandenberger,
L. H. Ford, A. Guth, A. Linde, F. C. Mena, V. F. Mukhanov, L. Parker, N. C. 
Tsamis and A. Vilenkin. We are also grateful to the University of Crete for 
its hospitality during portions of this project. This work was partially
supported by the Sonderforschungsbereich 375-95 f\"ur Astro-Teilchenphysik
der Deutschen Forschungsgemeinschaft, by DOE contract DE-FG02-97ER\-41029,
by NSF grant 94092715 and by the Institute for Fundamental Theory.

\end{document}